\documentclass[twoside,twocolumn,9pt]{article}
\usepackage{extsizes}
\usepackage[super,sort&compress,comma]{natbib} 
\usepackage[version=3]{mhchem}
\usepackage[left=1.5cm, right=1.5cm, top=1.785cm, bottom=2.0cm]{geometry}
\usepackage{balance}
\usepackage{mathptmx}
\usepackage{sectsty}
\usepackage{graphicx} 
\usepackage{lastpage}
\usepackage[format=plain,justification=justified,singlelinecheck=false,font={stretch=1.125,small,sf},labelfont=bf,labelsep=space]{caption}
\usepackage{float}
\usepackage{fancyhdr}
\usepackage{fnpos}
\usepackage[english]{babel}
\addto{\captionsenglish}{%
 
}
\usepackage{array}
\usepackage{droidsans}
\usepackage{charter}
\usepackage[T1]{fontenc}
\usepackage[usenames,dvipsnames]{xcolor}
\usepackage{setspace}
\usepackage[compact]{titlesec}
\usepackage{hyperref}

\usepackage{epstopdf}

\definecolor{cream}{RGB}{222,217,201}

\begin{document}

\pagestyle{fancy}
\thispagestyle{plain}
\fancypagestyle{plain}{
\renewcommand{\headrulewidth}{0pt}
}

\makeFNbottom
\makeatletter
\renewcommand\LARGE{\@setfontsize\LARGE{15pt}{17}}
\renewcommand\Large{\@setfontsize\Large{12pt}{14}}
\renewcommand\large{\@setfontsize\large{10pt}{12}}
\renewcommand\footnotesize{\@setfontsize\footnotesize{7pt}{10}}
\makeatother

\renewcommand{\thefootnote}{\fnsymbol{footnote}}
\renewcommand\footnoterule{\vspace*{1pt}%
\color{cream}\hrule width 3.5in height 0.4pt \color{black}\vspace*{5pt}} 
\setcounter{secnumdepth}{5}

\makeatletter 
\renewcommand\@biblabel[1]{#1}      
\renewcommand\@makefntext[1]%
{\noindent\makebox[0pt][r]{\@thefnmark\,}#1}
\makeatother 
\renewcommand{\figurename}{\small{Fig.}~}
\sectionfont{\sffamily\Large}
\subsectionfont{\normalsize}
\subsubsectionfont{\bf}
\setstretch{1.125} 
\setlength{\skip\footins}{0.8cm}
\setlength{\footnotesep}{0.25cm}
\setlength{\jot}{10pt}
\titlespacing*{\section}{0pt}{4pt}{4pt}
\titlespacing*{\subsection}{0pt}{15pt}{1pt}

\fancyfoot{}
\fancyfoot[LO,RE]{\vspace{-7.1pt}\includegraphics[height=9pt]{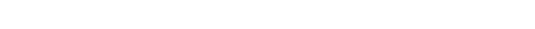}}
\fancyfoot[CO]{\vspace{-7.1pt}\hspace{13.2cm}\includegraphics{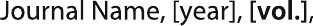}}
\fancyfoot[CE]{\vspace{-7.2pt}\hspace{-14.2cm}\includegraphics{head_foot/RF}}
\fancyfoot[RO]{\footnotesize{\sffamily{1--\pageref{LastPage} ~\textbar \hspace{2pt}\thepage}}}
\fancyfoot[LE]{\footnotesize{\sffamily{\thepage~\textbar\hspace{3.45cm} 1--\pageref{LastPage}}}}
\fancyhead{}
\renewcommand{\headrulewidth}{0pt} 
\renewcommand{\footrulewidth}{0pt}
\setlength{\arrayrulewidth}{1pt}
\setlength{\columnsep}{6.5mm}
\setlength\bibsep{1pt}

\makeatletter 
\newlength{\figrulesep} 
\setlength{\figrulesep}{0.5\textfloatsep} 

\newcommand{\topfigrule}{\vspace*{-1pt}%
\noindent{\color{cream}\rule[-\figrulesep]{\columnwidth}{1.5pt}} }

\newcommand{\botfigrule}{\vspace*{-2pt}%
\noindent{\color{cream}\rule[\figrulesep]{\columnwidth}{1.5pt}} }

\newcommand{\dblfigrule}{\vspace*{-1pt}%
\noindent{\color{cream}\rule[-\figrulesep]{\textwidth}{1.5pt}} }

\newcommand{\rkc}[1]{\textcolor{red}{(RKC: #1)}}
\newcommand{\xx}{\textcolor{red}{\large{XX}}}
\newcommand{\etal}{\textit{et al.~}}

\makeatother

\twocolumn[
 \begin{@twocolumnfalse}
{\includegraphics[height=30pt]{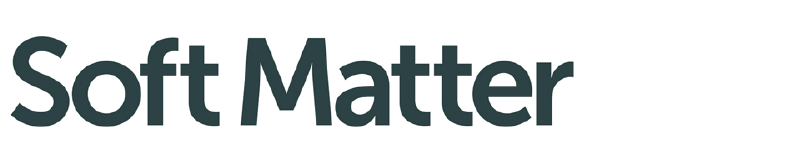}\hfill\raisebox{0pt}[0pt][0pt]{\includegraphics[height=55pt]{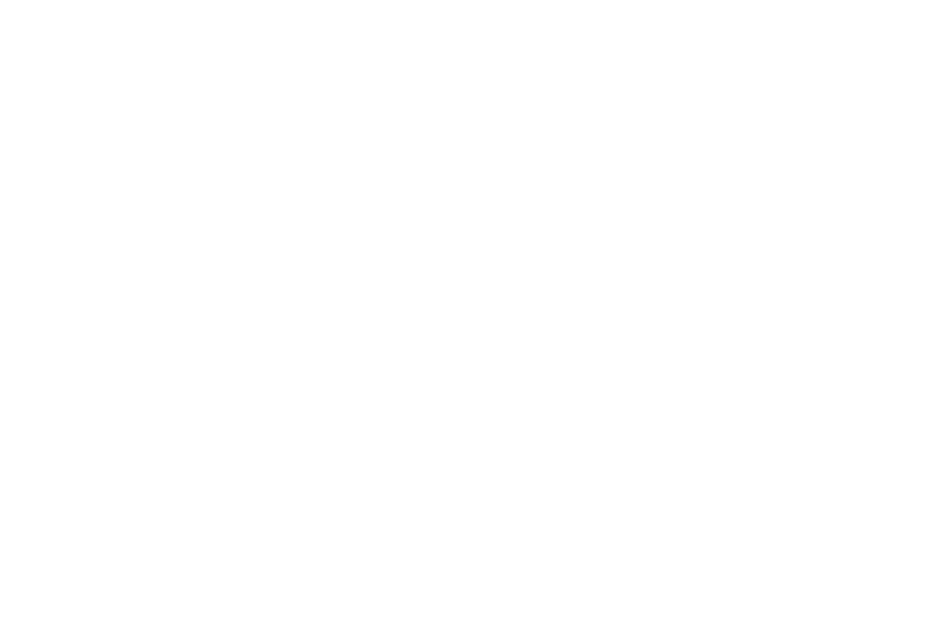}}\\[1ex]
\includegraphics[width=18.5cm]{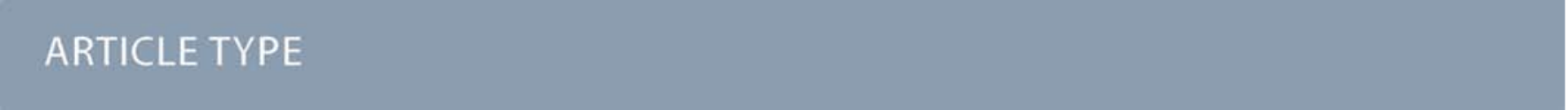}}\par
\vspace{1em}
\sffamily
\begin{tabular}{m{4.5cm} p{13.5cm} }

\includegraphics{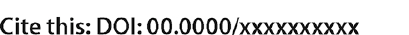} & \noindent\LARGE{\textbf{A Route to Hierarchical Assembly of Colloidal Diamond$^\dag$}} \\
\vspace{0.3cm} & \vspace{0.3cm} \\

 & \noindent\large{Yuan Zhou,\textit{$^{a}$} Rose K. Cersonsky,\textit{$^{bc}$} and Sharon C. Glotzer$^{\ast}$\textit{$^{abd}$}} \\

\textit{Submitted to Soft Matter\hspace{2in} October 3, 2021}  & \noindent\normalsize{Photonic crystals, appealing for their ability to control light, are constructed by periodic regions of different dielectric constants. Yet, the structural holy grail in photonic materials, diamond, remains challenging to synthesize at the colloidal length scale. Here we explore new ways to assemble diamond using modified gyrobifastigial (mGBF) nanoparticles, a shape that resembles two anti-aligned triangular prisms. We investigate the parameter space that leads to the self-assembly of diamond, and we compare the likelihood of defects in diamond self-assembled via mGBF \textit{vs.} the nanoparticle shape that is the current focus for assembling diamond, the truncated tetrahedra. We introduce a potential route for realizing mGBF particles by dimerizing triangular prisms using attractive patches, and we report the impact of this superstructure on the photonic properties.}

\end{tabular}

 \end{@twocolumnfalse} \vspace{0.6cm}

 ]

\renewcommand*\rmdefault{bch}\normalfont\upshape
\rmfamily
\section*{}
\vspace{-1cm}


\footnotetext{\textit{$^{a}$~Department of Chemical Engineering, University of Michigan, Ann Arbor, Michigan 48109, United States}}
\footnotetext{\textit{$^{b}$~Macromolecular Science and Engineering Program, University of Michigan, Ann Arbor, Michigan 48109, United States}}
\footnotetext{\textit{$^{c}$~Laboratory of Computational Science and Modelling, STI, \`Ecole Polytechnique F\`ed\`erale de Lausanne, 1015 Lausanne, Switzerland}}
\footnotetext{\textit{$^{d}$~Biointerfaces Institute, University of Michigan, Ann Arbor, Michigan 48109, United States, E-mail: sglotzer@umich.edu}}
\footnotetext{\dag~Electronic Supplementary Information (ESI) available: [details of any supplementary information available should be included here]. See DOI: 10.1039/cXsm00000x/}




\section{\textbf{Introduction}}
Photonic crystals are periodic patterns of materials with different dielectric constants; this microscopic patterning may lead to a photonic band gap, a region of frequencies that cannot propagate in the crystal, analogous to the electric band gap in semiconductors and insulators. When photonic band gaps occur in the visible spectrum, they can result in structural color, leading to the beautiful colors in the wings of \textit{Morpho rhetenor}\cite{kinoshita2002mechanisms, vukusic1999quantified}, the skin of the \textit{Furcifer pardalis} chameleon \cite{teyssier2015photonic}, and the peacock fern\cite{lee1997iridescent}. 

The first scientific observation of structural color was reported by \citet{Rayleigh1888} when he observed a change in materials coloration based upon the incidence angle of incoming light. A century later, \citet{Yablonovitch1987} gave photonic crystals their name and theorized that a vast array of photonic band gap crystals (PBGCs) existed. \citet{Ho1990} reported the first structure to exhibit a complete photonic band gap; the diamond structure, naturally found in Group IV elements, was capable of selective transmission of light when fabricated with spheres of high dielectric constant. Not only was diamond the first PBGC, for years, it has also remained the archetype for photonic crystals design\cite{Bragg1913, Maldovan2004, Cersonsky2018, cersonsky2021, Ducrot2017, Zhang2005, Wang, Liu2016, Damasceno2012d, Zanjani2016}. However, fabricating a colloidal diamond crystal has proven fraught with experimental difficulties. Defects in a photonic crystal can diminish the band gap by introducing new electromagnetic modes that propagate the previously reflected frequencies\cite{ashcroft1976solid}. Additionally, many popular self-assembly methods for isotropic colloids, including sedimentation, pressing, and centrifugation, are not suitable for open structures (\textit{i.e.} where spheres can only fit at very low filling fractions)\cite{garcia2001opal}. Radiation pressure or atomic force microscopy can trap and manipulate single particles and have led to diamond via etching a body center cubic(BCC) crystal\cite{garcia2002nanorobotic}; however, this process is costly and generally unscalable\cite{vogel2015advances}. Colloidal diamond using titania has also been synthesized using biological templating via a two-step templating process on Weevil scales;\cite{galusha2010diamond} however, this is limited by defects in the biologic templates themselves and raises bioethical and feasibility concerns for large-scale manufacturing.  In 2018, \citet{he2020colloidal} succeeded to self-assemble cubic diamond from tetrahedral clusters of partially-compressed spheres, with retracted sticky patches and a steric interlock mechanism that selects for the required staggered bond orientation. 

\begin{figure*}[h!]
\centering
\includegraphics[width=\linewidth]{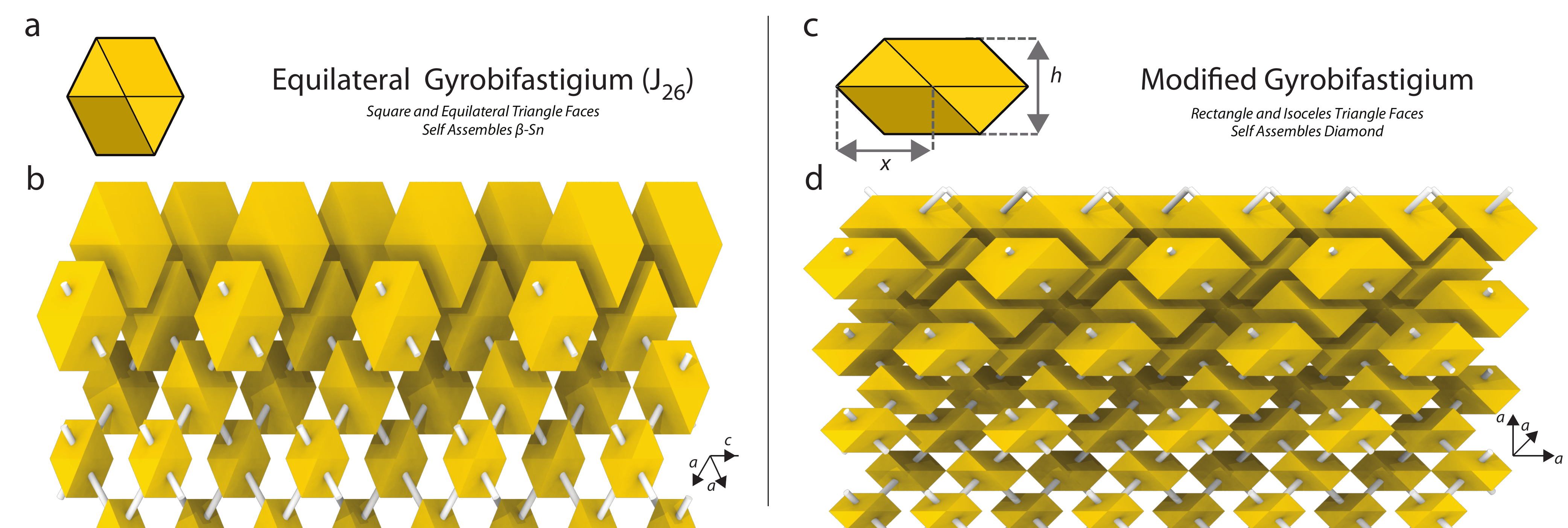}
\caption{\textbf{The eGBF and mGBF and their assembled structures.} (a) Gyrobifastigium with equal edges (eGBF), known as Johnson Solid 26 ($J_{26}$). (b) eGBF self-assembles $\beta$-Sn, shown here with diminishing particle size lower on the image for visibility. (c) The modified GBF (mGBF), related to the eGBF by uniaxial scaling. (d) The mGBF in the diamond structure. }
\label{fig:mGBF}
\end{figure*}

Polyhedrally shaped nanoparticles are a promising route for colloidal diamond, with entropy motivating face-to-face alignment and leading to self-assembly of complex crystals\cite{Damasceno2012}. In 2012, \citet{Damasceno2012d} reported an exciting prospect for diamond synthesis; truncated tetrahedral nanoparticles that self-assemble the diamond structure at packing fractions of ~0.6. This process has not yet led to the reliable fabrication of a diamond colloidal crystal \cite{nagaokaNature, sacannadiamond}. Furthermore, small changes in the truncation of the tetrahedra or increased pressure result in a lower-symmetry derivative \cite{Cersonsky2018} and slip planes often lead to hexagonal diamond domains when assembling via sedimentation \cite{gong2019entropy}.

Here, we propose an alternative nanoparticle shape that achieves a topologically protected \cite{zygmunt2019topological} diamond structure on the colloidal length scale: the modified gyrobifastigium(mGBF). The equilateral gyrobifastigium (eGBF) is an 8-sided polyhedron whose name means "two anti-aligned roofs" and is analogous to two regular triangular prisms misaligned by a 90$^\circ$ rotation and connected on their square faces. The eGBF was reported by \citet{Damasceno2012} and by \citet{sharma2018nucleus} to self-assemble $\beta$-Sn, a tetragonal relative of the diamond structure. Here, we report the assembly behavior of modified gyrobifastigia and demonstrate that it may be a better candidate than the truncated tetrahedron for nanoparticle self-assembly of colloidal diamond crystals. 

To date, gyrobifastigial nanoparticles have not been synthesized, although we suspect that this is more due to lack of incentive rather than difficulty. Given the literature on nanoparticles synthesis, there is good evidence to believe that one of the simplest routes to achieving a gyrobifastigium is through dimerizing anti-aligned triangular prism nanoparticles, as have been synthesized by Zhang \etal in 2010\cite{zhang2010photomediated}. We demonstrate the possibility of hierarchically assembling diamond by first assembling mGBFs as dimers from triangular prisms. We also report the effects of this superstructure on the photonic band gaps, as core-shell dimers would lead to a ``dimer diamond derivative'', or DDD, wherein two diamond lattices interpenetrate along the y-axis. 

\section{\textbf{Model and Methods}}

\subsection{Hard Particle Monte Carlo and Phase Determination}
In the canonical ensemble, wherein the number of particles ($N$), system volume ($V$) and temperature ($T$) are constant, the Helmholtz free energy is $A = U - TS$, where $U$ is the internal energy and $S$ is the entropy. For hard particles, $U$ is taken as $\infty$ where particles are overlapping, and 0 otherwise.

As each microstate is equally probable, per the Boltzmann equation, the entropy $S$ is proportional to the natural log of the number of microstates. From the second law of thermodynamics, the lowest free energy state corresponds to that which maximizes the entropy. In crowded systems of anisotropic particles, this leads to the self-assembly of ordered structures due to the entropy gained from face-to-face alignment\cite{Frenkel1999, Haji-Akbari2009, Agarwal2011, Damasceno2012d, Damasceno2012, Henzie2012,dijkstrasuperballs, dijkstratcube, VanAnders2014c, marson2014phase, VanAnders2014c, VanAnders2014d, Millan2014d, glotzer2015assembly, Schultz2015, Damasceno2015b, Harper2015a, harper2019entropic, Karas2019}.

We first conducted simulations to determine the packing fractions and aspect ratios which resulted in stable diamond assemblies. We ran separate NVT simulations  under periodic boundary conditions consisting of 5832 mGBF particles initialized in diamond at packing fraction $\phi \in {0.4-0.8}$ for different aspect ratios $h/x \in [0.35, \sqrt{2}]$. Each simulation was run for $1.5\times10^6$ MC sweeps or until the structure melted. We omitted aspect ratios that are geometrically forbidden in diamond at the target $\phi$, as noted in the phase diagram. For each set of simulation parameters, we ran three independently-seeded replicas. For those sets of parameters that retained the diamond structure, We then ran three independently-seeded $NVT$ simulations of 512 mGBF particles initialized in a low-density fluid. The simulation boxes were compressed isotropically from the fluid phase to the lowest packing fraction at which diamond is stable. These simulations were each run for $5\times10^7$ Monte Carlo (MC) sweeps. All simulations used the open-source molecular dynamics package HOOMD-Blue\cite{anderson2019hoomd, Anderson2016}. The computational workflow and data management were supported by the \textit{signac} and \textit{signac-flow} frameworks\cite{Adorf2018, ramasubramani2018}.

Using \textit{freud}\cite{Ramasubramani2020}, and \textit{scikit-image} \cite{ScikitImage}, we identified the self-assembled crystals by comparing the bond-order diagrams (BODs) of the first neighbor shell with a library of common self-assembled structures at the same packing fraction. We computed bond order diagrams in spherical coordinates across $\varphi$ and $\theta$ with a 100x100 grid. We then identified a confidence score based upon the structural similarity index (SSI) of the BODs and reference structures and denoted by the opacity of the phase diagram. The reference structures here used are fluid, diamond, tetragonal diamond\cite{Cersonsky2018}, $\beta$-Sn, face-centered cubic (\textit{i.e.} fcc or cubic close-packed), body-centered cubic (bcc), and simple cubic (sc)
We visualize our crystals in Fig.~\ref{fig:mGBF} using \textit{OVITO}\cite{stukowski2009visualization}.

\begin{figure*}[h!]
\centering
\includegraphics[width=\linewidth]{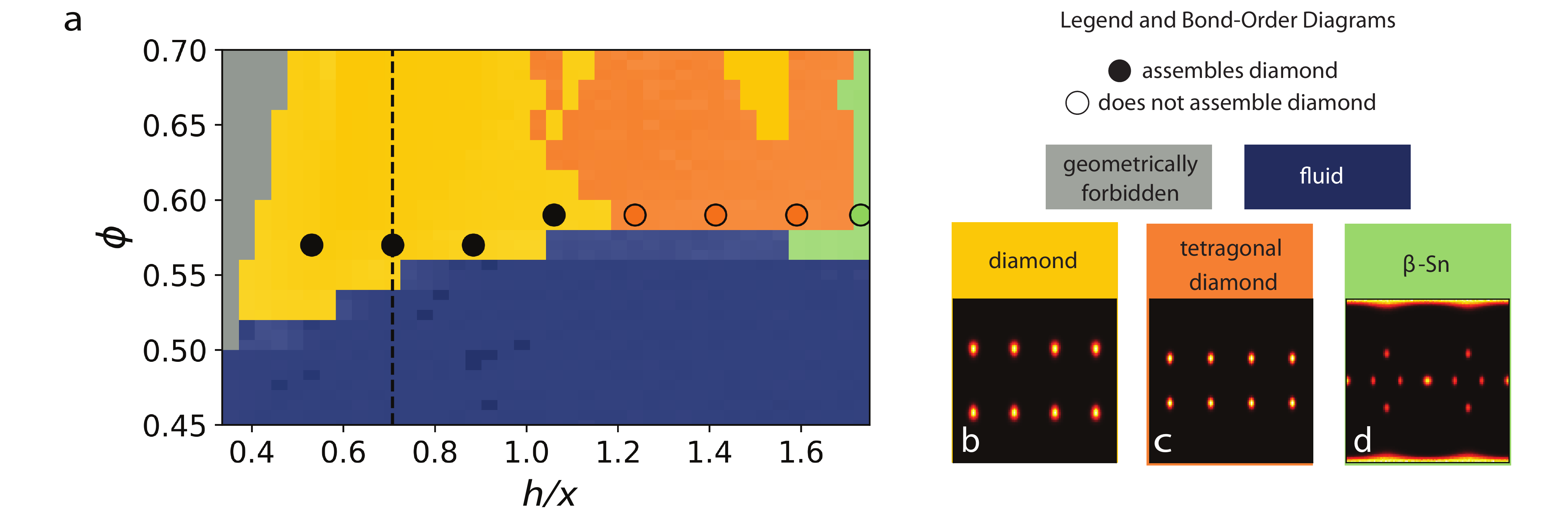}
\caption{\textbf{Phase diagram with aspect ratio.} (a) The phase diagram of mGBF at different aspect ratios ($h/x$) found through melting simulations for different packing fractions $\phi$. The region in yellow shows a wide range of ($h/x$, $\phi$) for which the cubic diamond structure is stable; here, the dashed line denotes $h/x=\sqrt{1/2}$, where the particles form a perfect $109.5^\circ$ bond angle when aligned. As $h/x$ increases, tetragonal diamond (orange region) and $\beta$-Sn (light green region) form, where the eGBF transitions to $\beta$-Sn as expected. Opacity shows the square of the SSI, \textit{i.e.} confidence in identifying the structures, with the minimum confidence level of 0.7332. Points marked as circles denote the systems where self-assembly simulations were also run, with systems assembling the cubic diamond structure denoted by solid symbols and those that failed to assemble the cubic diamond structure by open symbols. (b-d) Reference bond order diagrams for cubic diamond, tetragonal diamond, and $\beta$-Sn, respectively.}
\label{fig:phase_diagram}
\end{figure*}

\subsection{Design of Patchy Particles to Assemble Dimer Diamond Derivative (DDD)}

The mGBF is isomorphic to conjoined anti-aligned triangular prisms, giving the possibility to create the mGBF through dimerization. Using the JIT\cite{anderson2019hoomd} plug-in to HOOMD-Blue, we simulated triangular prisms with attractive patches modeled by a square well (SW) potential $U\left(r_{ij}\right) = -\epsilon$ for $0 < r_{ij} \leq \sigma$, where $\epsilon$ is the strength of the attraction (in units of $kT$) and $\sigma$ denotes the attractive distance (in units of particle size). The positions of patches were placed diagonally on the largest square facet to encourage the 90$^\circ$ misalignment using a single particle type. We parameterize the locations of the patches using the distance from the center of the bottom face, where $d=0$ denotes patches at the center of the face and $d=1$ denotes patches at opposing vertices.

We ran simulations of 5832 dimers for $d = 0.2, 0.6, 0.8, 1.0$ within a parameter space defined by $\epsilon \in \{0.05, 0.1, 0.5, 1.0\}$, and $\sigma \in [0.05, 0.25]$, $d\sigma = 0.05$ at packing fraction $\phi = 0.60$, to find the optimized SW potential parameters for phase stability\footnote{We omit d=0.4 for its likeness to d=0.2}. The SSI is used to be the evaluation reference of the degree of melting during the stability test simulation.

For each patch location, 1000 dimers were initialized in DDD at $\phi = 0.60$ with the optimized SW potential parameters. We then expanded simulation boxes isotropically to determine the melting packing fraction $\phi_{melt}$. These simulations occurred over $1.5 \times 10^6$ MC sweeps. Then, at packing fraction $\phi = 0.60$, we conducted assembly simulations of 1024 dimers of the same optimized potential but different patch locations with a pre-formed diamond seed containing 128 triangular prisms. These simulations were accelerated by performing parallel runs using MPI domain decomposition on multiple central processing units (CPUs) for a total of $8\times10^7$ MC sweeps\cite{Glaser2015}. 

\subsection{Photonic Band Structure Calculations}
To decouple the length scales at which self-assembly occurs ($\phi \geq 0.55$) from those at which photonic band gaps occur ($\Phi \leq 0.4$), we assume each mGBF particle to be composed of different materials in a core-shell design, similar to \citet{Cersonsky2018}. For a ``direct'' structure, this entails a high dielectric core and low dielectric shell, with the reverse for the ``inverse'' structure. To explore the effects of this substructure on the photonic properties, we use MIT Photonic Bands (MPB)\cite{Johnson2001}, a software that computes the eigenmodes of Maxwell's equations in periodic dielectric structures for using a planewave basis.

We computed the photonic band structure for a range of core radii and dielectric constants for direct and inverse structures. For all computations, we used a mesh size of 3, k-point interpolation of 10, and computed the first 100 photonic bands across the 1st Brillouin zone (determined by \textit{spglib}\cite{spglib}). Consistent with convention, we normalized the photonic band gaps to be dimensionless, $\Delta\omega/\omega^*$, where $\Delta\omega$ is the width of the complete gap window, and $\omega^*$ is the mid-gap frequency. The frequency used here has units of (speed of light/$a$) where $a$ is the lattice constant.

\section{\textbf{Results}}

\subsection{Phase Diagram of mGBF}

The eGBF assembles $\beta$-Sn ($tI$4, $I 4_1/amd$), which is related to diamond ($cF$8, $Fd\bar{3}m$) through a uniaxial scaling (as shown in Fig.~\ref{fig:mGBF}). We propose to modify the shape of the eGBF by a uniaxial scaling of the relative height $h/x$. The height of the equilateral GBF is approximately 2 times its lateral dimensions $h/x=\sqrt{3}/2\approx1.73$. Thus, reducing the difference in these dimensions, \textit{i.e.} reducing the aspect ratio (denoted in Fig.~\ref{fig:mGBF} on the modified GBF), may result in a cubic structure. This points to $h/x=1/\sqrt{2}\approx0.707$ as an ideal aspect ratio for a mGBF, where the aligned faces would coincide with a perfect 109.5$^\circ$ bond angle. This is denoted in Fig.~\ref{fig:phase_diagram}(a) with a black dotted line.

We surveyed a range from $h/x \in {0.25-1.75}$. We report the phase diagram (Fig.~\ref{fig:phase_diagram}a), displaying the phases for all melting simulations for varying packing fractions and $h/x$ ratios after $1.5\times10^6$ MC steps. BODs of reference structures, diamond (Fig.~\ref{fig:phase_diagram}b), tetragonal diamond (Fig.~\ref{fig:phase_diagram}c), $\beta$-Sn (Fig.~\ref{fig:phase_diagram}d) and fluid are used to identify the phase of systems after melt simulations. BODs of common entropic crystals fcc, bcc, and sc were also compared, but yielded no matches.

\begin{figure}
\centering
\includegraphics[width=\linewidth]{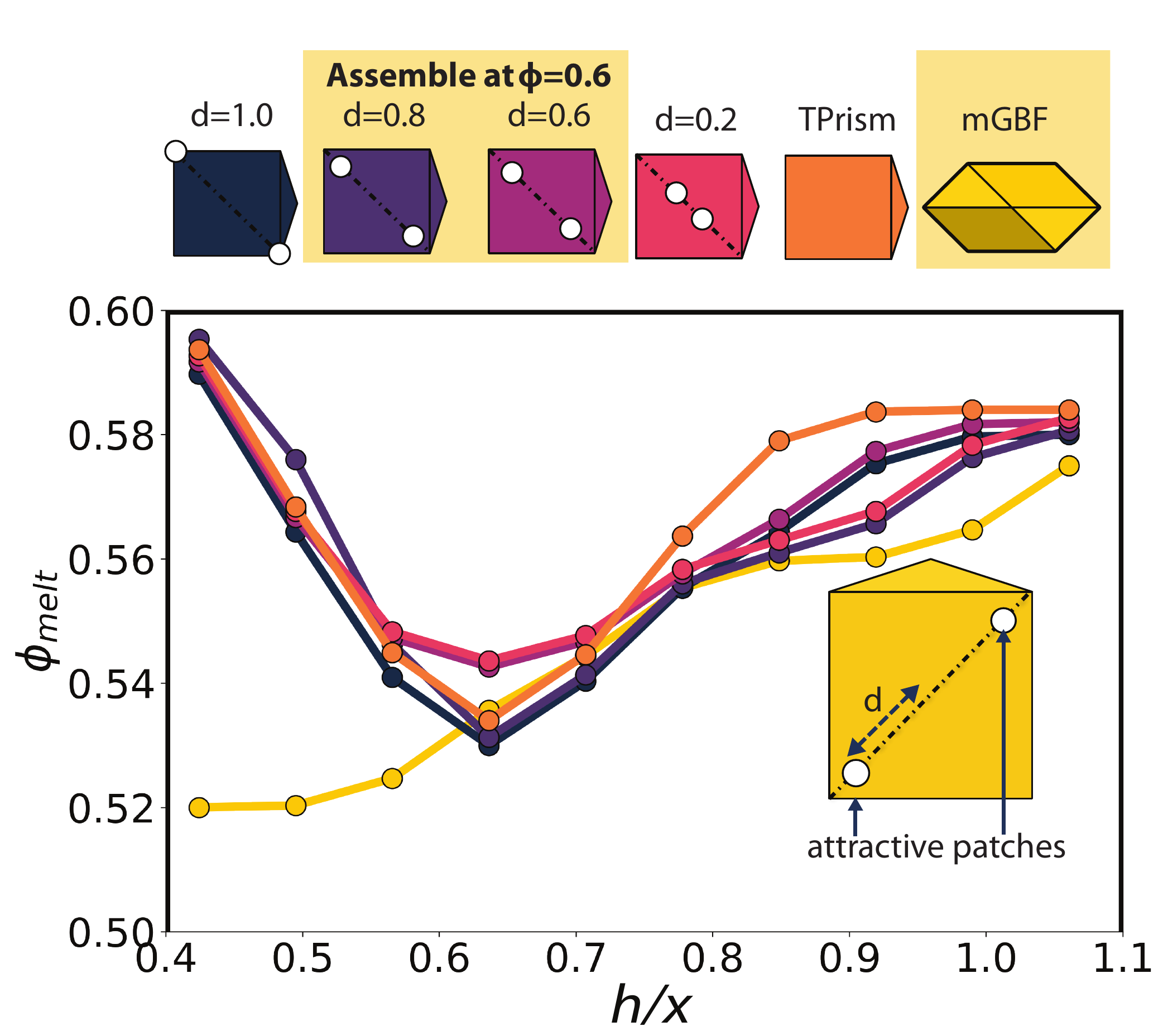}
\caption{\textbf{Stability of Patchy Dimers in Diamond.} The melting points of diamond formed by mGBF, DDD formed by both interacting and non-interacting triangular prisms (TPrism) for different prism height ratios and patch locations. Here, the optimized SW parameters of the dimers are $\sigma=0.1$, $\epsilon=0.1$ and d=0.8. Particles with a yellow background assemble cubic diamond in separate simulations at $\phi=0.60$. (Inset). The design pattern of patchy triangular prisms. The attractive patches are placed diagonally to promote anti-alignment using a single-component system. The attraction follows the formula of square-well potential, including parameters attractive distance $\sigma$ and attractive strength $\epsilon$. $d$ here denotes the distance between the patch to the center of the bottom facet. }
\label{fig:patch}
\end{figure}

Systematic study indicates the SW potential parameters $\epsilon=0.1$ and $\sigma=0.1$ are sufficient to stabilize the dimers in all patch configurations (see Fig.~\ref{fig:sw_param}). The locations of patches on the prism is a key factor in stabilizing DDD. In Fig.~\ref{fig:patch}, we show the melting packing fraction $\phi_{melt}$ for the mGBF diamond and dimer-assembled diamond derivative for a range of patch locations and mGBF aspect ratios $h/x \in (0.4,1.1)$. SSI was used to evaluate the phase behavior and stability of DDD formed by the dimers with and without patches, as shown in Fig.~\ref{fig:patch}. Here we observe that for $h/x = 0.6 - 0.8$, the dimers melt at similar packing fractions to the mGBF, with melting occurring at higher $\phi$ for $h/x < 0.6$ and $h/x >0.8$. 

We then attempted self-assembly from fluid using $h/x=0.6363$ (where the minimum in melting packing fraction occurs) and $\phi = 0.6$, which was successful for intermediate $d = 0.6, 0.8$. Prisms with patches located at $d=1.0$ failed to crystallize due to the randomness of interactions between vertices, whereas for $d< 0.6$, the additional degrees of rotational freedom impeded self-assembly.

\subsection{Effects on Photonic Properties}

\begin{figure}[h!]
\centering
\includegraphics[width=\linewidth]{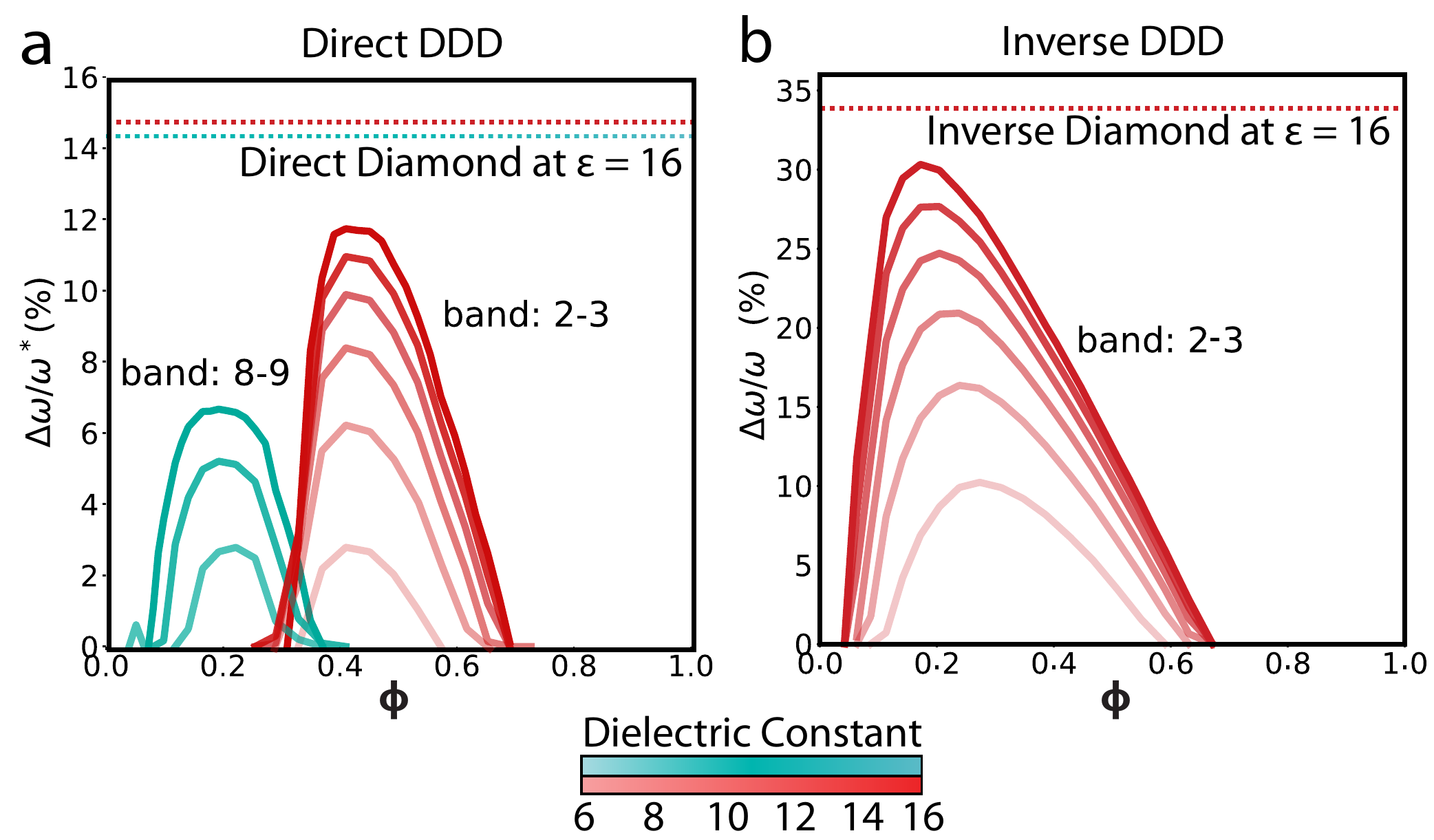}
\caption{\textbf{Photonic properties of dimer diamond derivative (DDD) in (a) direct and (b) inverse forms.} Gap sizes ($\Delta\omega/\omega^*$) are given with respect to the filling fraction ($\Phi$) of dielectric cores for dielectric contrasts of $\varepsilon = 6, 8, 10, 12, 12, 14, 16$. Dashed lines show the maximum gap sizes for diamond structure across all $\Phi$ at $\varepsilon=16$. Different colors are used to represent the location of the gaps, with teal for gaps between bands 8-9 and red for gaps between bands 2-3. Opacity of lines are for different dielectric constants shown in the figure legend, where the higher dielectric constant is more opaque.}
\label{fig:dimer-photonic}
\end{figure}

With polyhedral nanoparticles, it is necessary decouple the length scale of the nanoparticles from that of the high dielectric medium, suggesting a core-shell model. With this design strategy, there are two possibilities for placing the high and low dielectric material -- either the monomers join to have one common core or have individual high or low dielectric medium cores, leading to the distinct DDD. However, given the earlier focus on photonic band gaps, this requires additional calculations to ensure that DDD exhibits similar optical properties to diamond, as is the case with many diamond derivatives\cite{Cersonsky2018, cersonsky2021}.

Using MPB, we calculated the photonic band structures with different dielectric constants and filling fractions of higher dielectric medium $\Phi$ for both the direct and inverse structures, where the cores have the high or low dielectric constant, respectively. There are complete photonic gaps in direct diamond between bands 2 and 3, band 8 and 9, and 14 and 15, and for inverse diamond between 2 and 3. DDD has similar gaps between bands 2-3 and 8-9, and of proportional size, as noted in Fig.~\ref{fig:dimer-photonic}.

We investigated PBGs for different $\varepsilon \in (6,16)$--even with a low dielectric constant as $\varepsilon =6$, DDD has gaps up to $\approx 3\%$ in direct structures and $\approx 10\%$ in inverse structures. All gaps increase with greater dielectric contrast.

\subsection{Comparison with Truncated Tetrahedra}

\begin{figure}[h!]
\centering
\includegraphics[width=\linewidth]{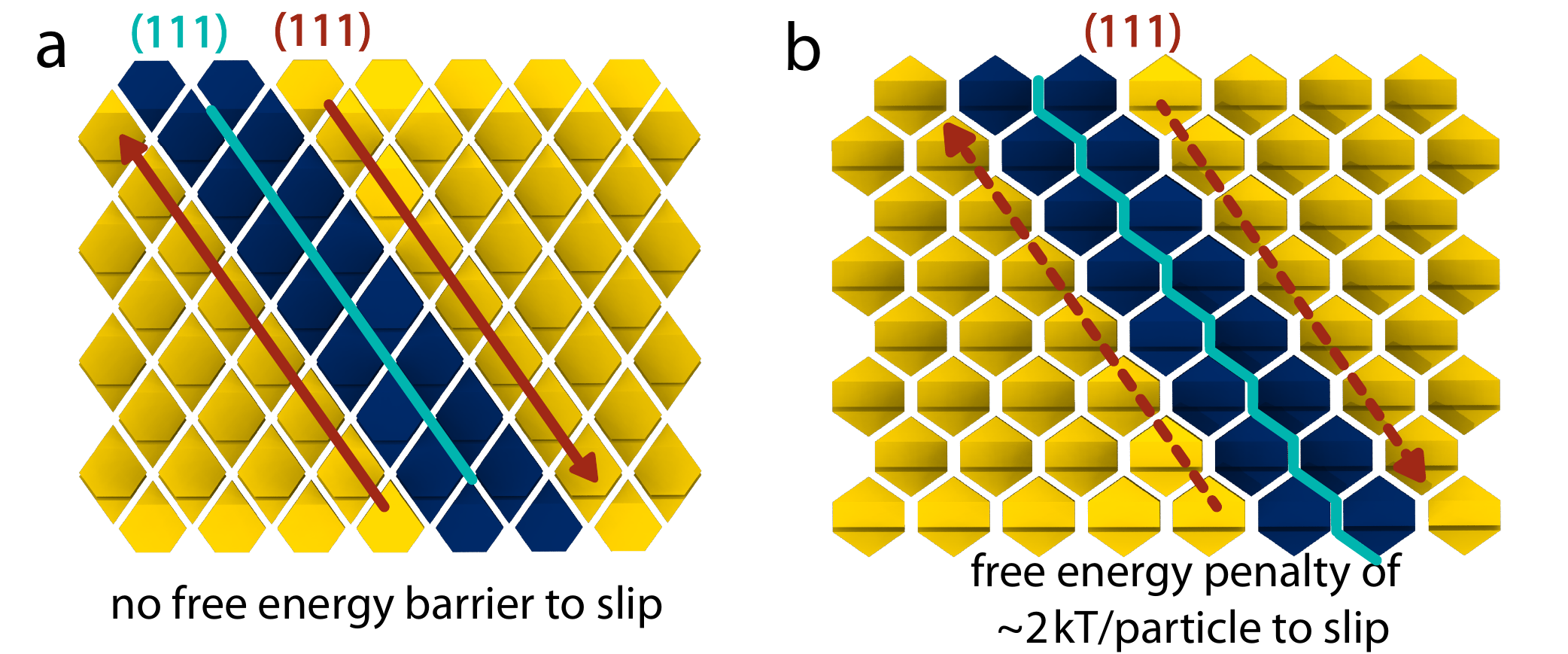}
\caption{\textbf{Slip plane in diamond structures assembled by truncated tetrahedra and mGBF.} A [10$\bar1$] cross-section of our diamond structures assembled by two different shapes. (a) The (111) slip planes (denoted by a teal line) in truncated tetrahedra-assembled diamond. The slip plane can move without free energy penalty. (b). The geometry of slip plane in mGBF-diamond. The triangle wave interdigitation impedes slip, helping to stabilize the crystal.}
\label{fig:slip_plane}
\end{figure}

\begin{figure}
\includegraphics[width=\linewidth]{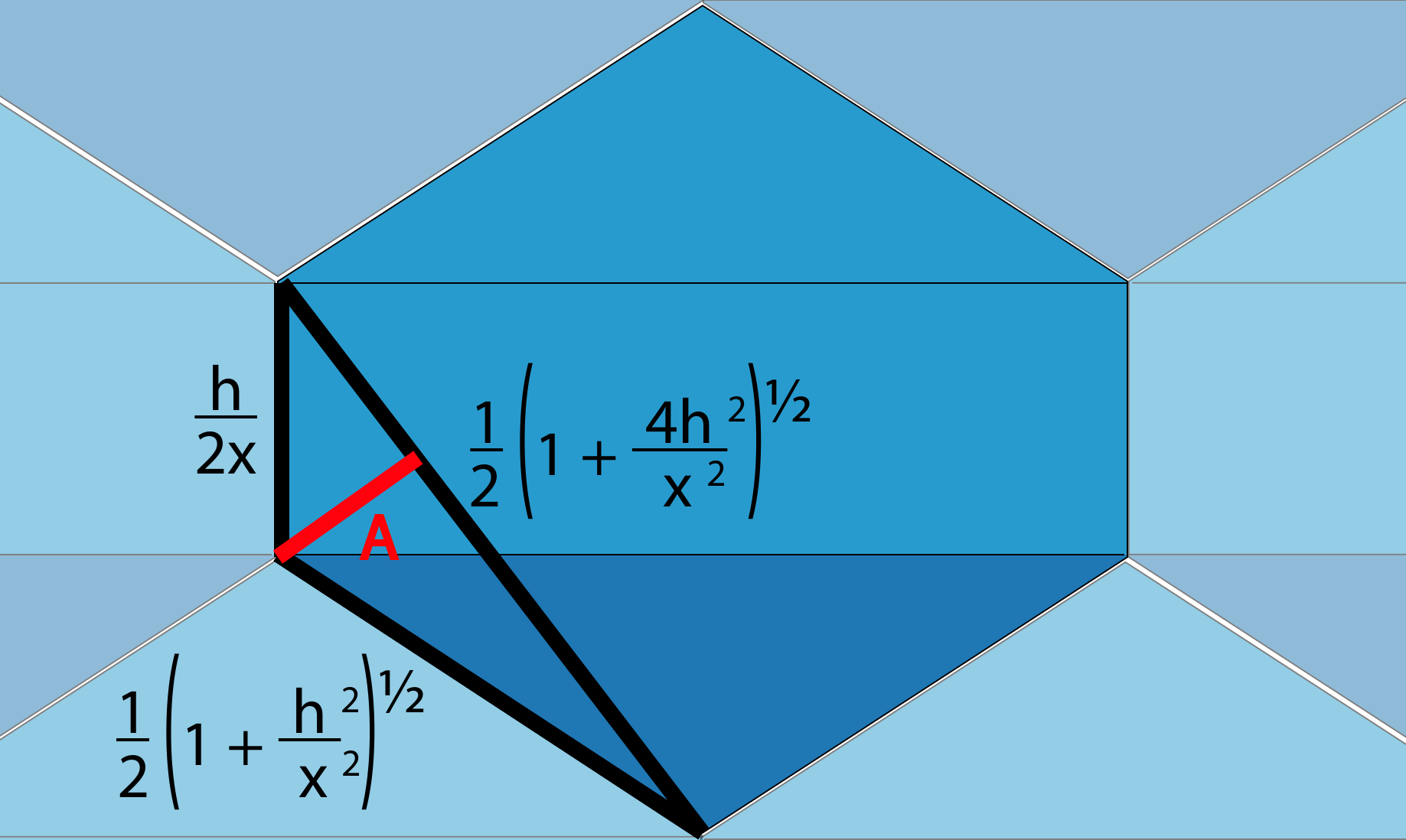}
\caption{\textbf{The mGBF diamond assembly through a shape allophile lens.} Considering the plane perpendicular to the $\{111\}$ slip, the allophilic amplitude is computed from the triangular lengths of the interdigitation. This impedes slip on the order of 2kT / particle, as estimated from \citet{Harper2015a}}
\label{fig:mgbfallo}
\end{figure}

In the truncated tetrahedron assembled diamond crystal, there are slip planes oriented along the $\{111\}$ direction, as highlighted in the [10$\bar1$] cross-sectional view in Fig.~\ref{fig:slip_plane}. However, in mGBF-assembled diamond, the interdigitation of the mGBF particles makes slip difficult, similar to shape allophiles\cite{Harper2015a} and lock-and-key colloids\cite{pacman}. Only line defects (dislocations) are possible along the [10$\bar1$] or [101] direction (depending on the orientation of the lines). From \citet{Harper2015a}, we can estimate, based upon $\phi = 0.60$ and $h/x = 1/\sqrt{2}$, that slip along the typical slip directions will incur a free energy penalty of approximately $2kT$ per particle. To see this, note that the mGBF interdigitates along the (111) plane, analogous to shape allophiles in 2D. Here we estimated the free energy barrier to slip along this plane using the data introduced by \citet{Harper2015a}. First, we assume the mGBF assembly is analogous to shape allophiles by only considering a 2D plane perpendicular to the slip direction, as visualized in Fig.~\ref{fig:mgbfallo}. Here, we assume $n_k = 1$ and determine an amplitude of the interdigitation as a function of the $h/x$ ratio, where 

\begin{equation}
A = \frac{(h/x)}{4 * (h/x) ^2 + 1}
\label{eq:amp}
\nonumber
\end{equation}

\noindent For $h/x \in [0.35, \sqrt{2}]$, this yields an amplitude $0.16 - 0.25$. From these values and Fig. S3-4 in \citet{Harper2015a}, the slip penalty is estimated to be on the order of ~2kT per particle in this 2D plane. These estimates suggest that the mGBF-assembled diamond will be more topologically protected, as discussed in \citet{zygmunt2019topological}, than that assembled by truncated tetrahedra and open the possibility of self-assembly via sedimentation, which is prohibited in truncated tetrahedra due to mis-stacking\cite{gong2019entropy}.

The dimensionality of defects in photonic crystals also determines the shapes and properties of the localized photonic states in the gap: microcavities occur with point defects, waveguides with line defects, and mirrors with planar defects\cite{joannopoulos1997photonic}. In general, line defects can be used to design complex photonic devices, including wavelength scale optical waveguides or resonators\cite{notomi2001extremely, lau2002creating, shinya2003single}, further supporting the pursuit of the mGBF-assembled diamond. Diamond crystals with line defects can also give rise to additional photonic phenomena, including interface states\cite{karathanos1994planar} and featureless and energy-dependently scaled absorption spectra\cite{hounsome2005optical}. 

\section{\textbf{Conclusions}}
We reported a promising route towards self-assembling a colloidal diamond crystal using the modified gyrobifastigium (mGBF) that provides topological protection. Furthermore, by hierarchically assembling the mGBF using triangular prisms dimers, we have introduced further opportunities for synthesis and customization.

\section*{Conflicts of interest}
There are no conflicts to declare.

\section*{Acknowledgements}
This material is based upon work supported by the Department of the Navy, Office of Naval Research under ONR award number N00014-18-1-2497. This research utilized
computational resources and services supported by Advanced Research Computing at the University of Michigan, Ann Arbor, and used the Extreme Science and Engineering Discovery Environment (XSEDE), which is supported by National Science Foundation Grant ACI-1053575 (XSEDE Award DMR 140129)\cite{xsede}. 
We thank B. Dice, P. Zhou, V. Ramasubramani, P. Lawton and J.A. Anderson for helpful discussions. 



\balance


\bibliography{rsc} 
\bibliographystyle{rsc} 

\appendix
\renewcommand\thefigure{\thesection.\arabic{figure}} 
\setcounter{figure}{0}      

\section{Optimization of the patch design}
\begin{figure}[h!]
\centering
\includegraphics[width=\linewidth]{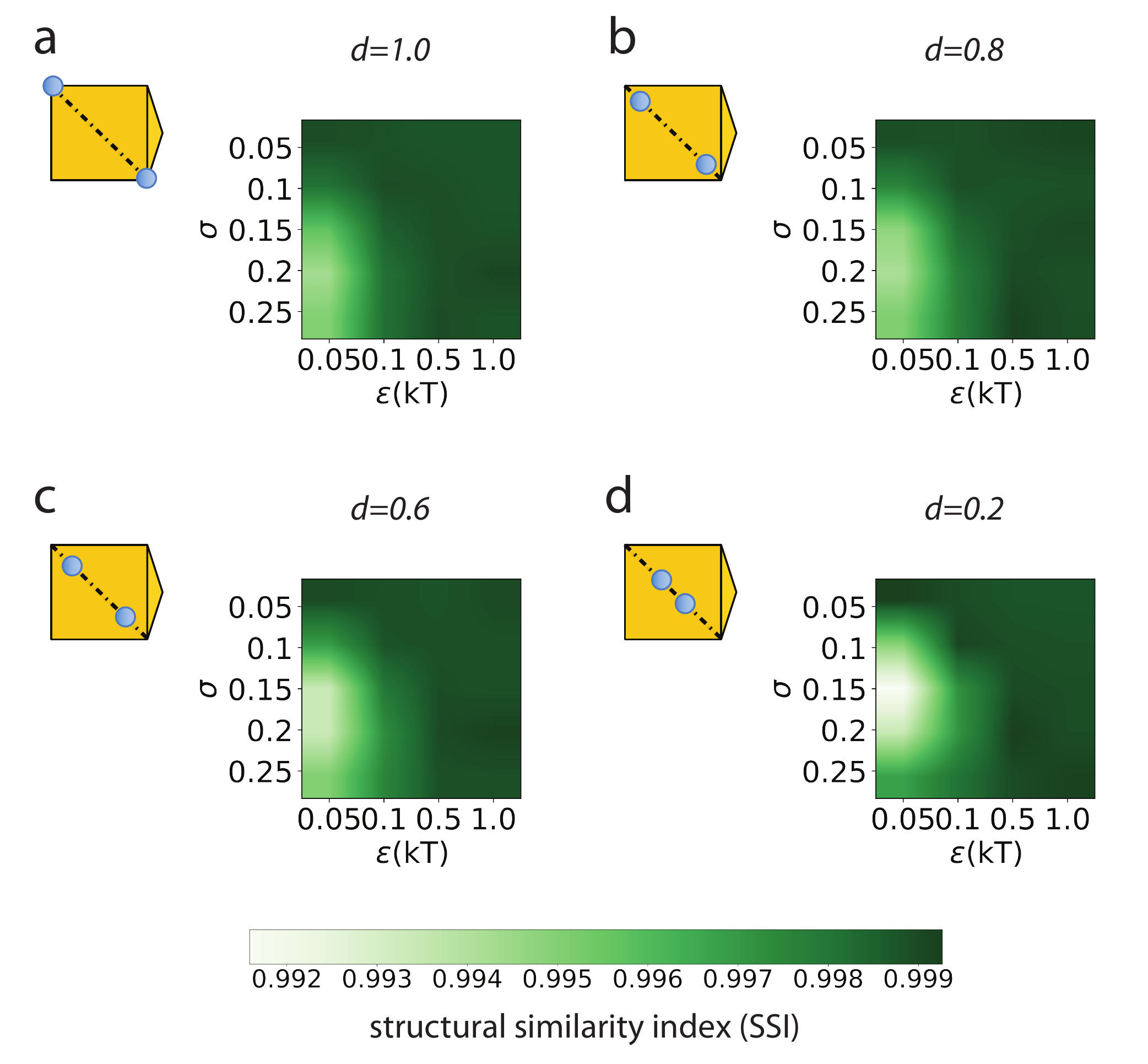}
\caption{\textbf{Stability of dimer-diamond with various patch conditions.} (a-d) Parameters of S-W potential with $\epsilon \in \{0.05,0.1,0.5,1.0\}$ and $\sigma \in \{0.25,0.2,0.15,0.1,0.05\}$ are studied under different distributions of patches ($d \in \{1.0,0.8,0.6,0.2\}$, as shown in inserts of (a-d)). Similarity between the BODs of simulations and reference used as the criterion of stability of the double diamond.}
\label{fig:sw_param}
\end{figure}

Parameter of square-well potential and the locations of patches are varied in a reasonable range, then optimized. The best parameter set chosen is {$\sigma=0.1$, $\epsilon=0.1$ and d=0.8} (Fig.~\ref{fig:sw_param}).

\section{Photonic Properties of DDD}
\begin{figure}[h!]
\centering
\includegraphics[width=\linewidth]{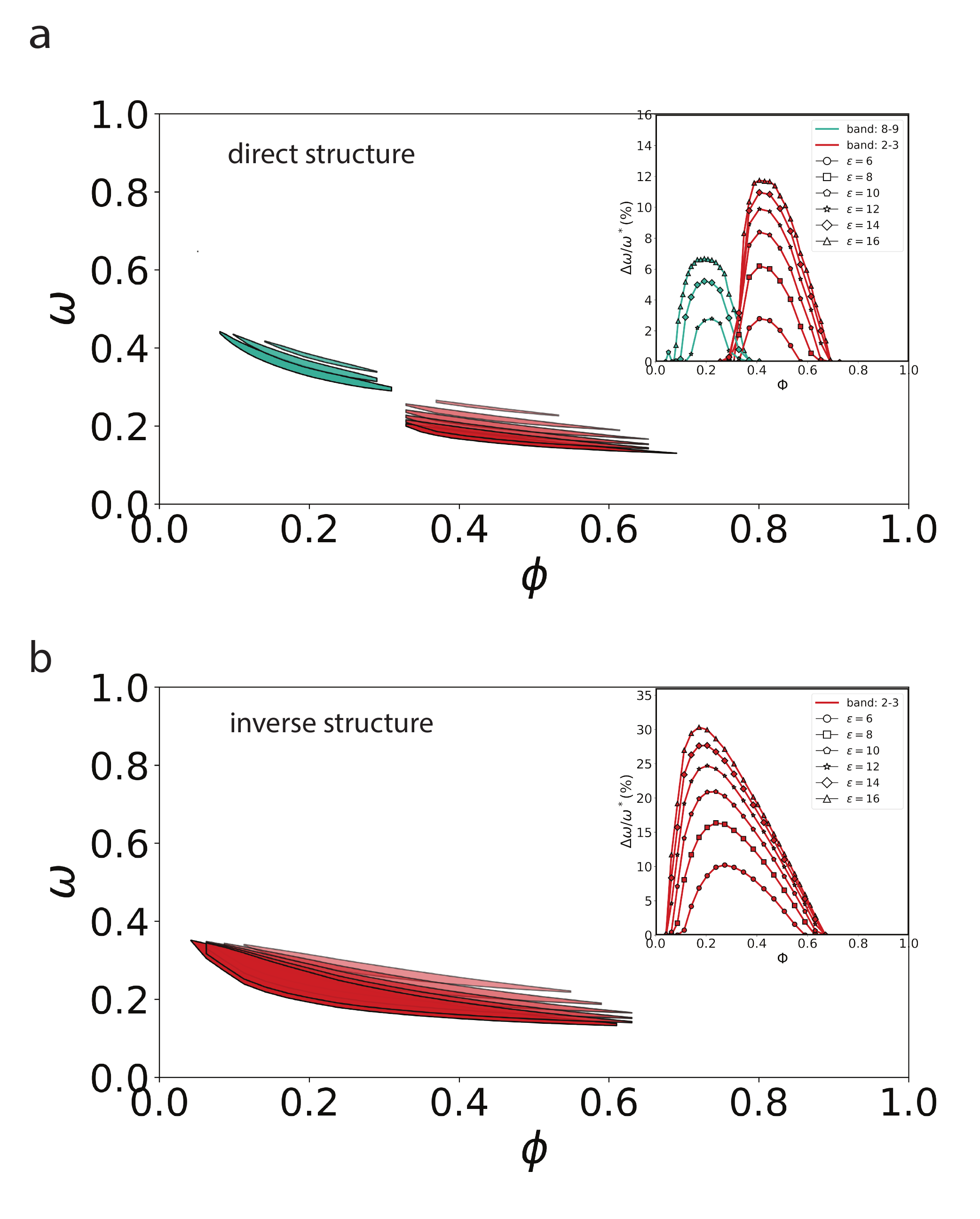}
\caption{\textbf{Gaps sizes of double diamond with different dielectric constants for both direct and inverse structures.}}
\label{fig:eps}
\end{figure}

Here we report the gap atlases of DDD, which are summarized in the main text and show in Fig.~\ref{fig:eps}. In each contour, opacity denotes the dielectric contrast, with more opaque contours corresponding to higher contrast. Dimensionless frequency $\omega$ is given in units (speed of light / $a$), where $a$ is the lattice parameter. To convert from dimensionless frequency, one can use the relation:

\begin{equation}
  \lambda=\frac{a}{\omega}
  \nonumber
\end{equation}
where $\lambda$ is the observable wavelength of the photonic band gap.

\end{document}